\documentstyle[twoside,fleqn,singularities]{article}

\newcommand{\noi}{\noindent}
\newcommand{\ra}{\rightarrow}
\newcommand{\bea}{\begin{eqnarray}}
\newcommand{\eea}{\end{eqnarray}}

\newcommand{\rw}{RW$\!\phi$}

\newcommand{\hold}{L^2(0,\infty)}
\newcommand{\hnew}{L^2({\bf R})}

\def\r{{\bf R}}

\newcommand{\AmS}{{\protect\the\textfont2
  A\kern-.1667em\lower.5ex\hbox{M}\kern-.125emS}}

\title{Some remarks on singularities in quantum cosmology\thanks{To
appear in: {\sl Proc. of the Second Conference on Con\-strain\-ed
Dynamics and Quantum Gravity,} Santa Mar\-gher\-ita Ligure,
Italy, 17-21 September 1996. Edited by V. de Alfaro et al. Nuc.
Phys. B (Proc. Suppl.), 1997.}}

\author{Mark J. Gotay\address{Department of Mathematics, University of
Hawai`i, Honolulu, HI 96822 USA}\thanks{Supported in part by NSF grant DMS
96-23083. Email: gotay@math.hawaii.edu}
        and
        Jacques Demaret\address{Institut d'Astrophysique,
Universit\'e de Li\`ege, B-4000 Li\`ege, Belgium}\thanks{Supported in part
by contract No. ARC 94/99-178
``Action de la Recherche Concert\'ee de la Communaut\'e Fran\c caise''
(Belgium). Email: demaret@astro.ulg.ac.be}}

\begin{document}

\begin{abstract}
We discuss to what extent classical singularities
persist upon quantization in two simple cosmological models.
\end{abstract}

\maketitle

\setcounter{footnote}{0}

\section{INTRODUCTION}

The question of whether classical singularities persist in quantum
cosmology remains a fascinating one. In \cite{gd}, we conjectured that
(F) {\sl self-adjoint quantum dynamics in a fast-time gauge is
singular\/}, whereas (S) {\sl self-adjoint quantum dynamics in a
slow-time gauge is always nonsingular\/}. By a ``fast-time gauge'' we
mean a choice of time $t$ such that the classical singularities occur
at $t = +\infty$ or $t = -\infty$. A time $t$ is ``slow'' if the
singularities occur when $|\,t\,| < \infty.$

In this paper, we verify Conjecture (F) for a $k = 0$ Robertson-Walker
cosmology containing a massless scalar field in a matter-time gauge.
We also discuss the status of Conjecture (S) in the case of a
dust-filled Friedmann-Robertson-Walker cosmology in an extrinsic-time
gauge. These two examples have drawn some attention recently
\cite{l90,l91,l96}, and are interesting as they exhibit certain
features which have not received adequate consideration in the
literature.

\section{RW$\!\phi$ MODELS}

We start with a  $k = 0$ Robertson-Walker cosmology filled with a
massless scalar field $\phi$. Detailed discussions of these
``RW$\!\phi$'' models can be found in \cite{l96,bi,gi1}; the
background we need can be  summarized as follows. Upon making an ADM
reduction by choosing the intrinsic-time gauge $t = \phi$, the
dynamics of this model can be described in terms of the canonical
variables  $(R,\pi_{\! R})$, where
$R > 0$ is the classical radius and $\pi_{\! R} \neq 0.$ (The reason
why $\pi_{\! R} \neq 0$ when $k \leq 0$ is explained in \cite{l96}.) The
phase space thus has two components,
$(0,\infty) \times (-\infty,0)$ and $(0,\infty) \times (0,\infty),$
and the Hamiltonian is
\begin{equation} H(R,\pi_{\! R}) =
\frac{1}{\sqrt{12}}\,R\,|\pi_{\! R}|.
\label{eq:ch}
\end{equation}

\noi When $\pi_{\! R} > 0$, the model has an initial singularity at
$t = -\infty$ and expands thereafter. The situation is time-reversed
when $\pi_{\! R} < 0$. Thus $t = \phi$ is fast.

The quantum dynamics of the RW$\!\phi$ models in various fast-time
gauges have been extensively studied
\cite{bi,gi1,gi2}. In all cases the quan\-tized models were found to be
singular. We will prove that the  quan\-tized $k = 0$ model with $t
= \phi$ is singular as well, contrary to the assertion in \cite{l96}.

\subsection{Canonical Quantization}

Canonically quantizing, we take the Hilbert space to be the orthogonal
direct sum $L^2(0,\infty) \oplus L^2(0,\infty)$, corresponding to the
two components of the phase space. The quantum Hamiltonian is
$\hat H = \hat H_- \oplus \hat H_+$ acting on the first and second
summands, respectively, where
\begin{equation}
\hat H_{\pm} = \mp \frac{i\hbar}{\sqrt{12}}\left(\!R\,\!\frac{d}{dR} +
\frac{1}{2}\right).
\label{eq:qh}
\end{equation}

\noi Note that as $H$ is linear in the momentum on each component, it
is (unambiguously) quantized according to the ``product $\ra$
anticommutator'' rule. Now $\hat H$ commutes with the projectors
onto each summand, whence the two halves of the Hilbert space are
dynamically decoupled, i.e., there can be no tunneling between
states in them. It therefore suffices to
study the quantum dynamics on each summand individually.\footnote{ See
\cite[\S6.4.3]{ta} for remarks on the quantum mechanics of
systems with disconnected phase spaces.} We will
accordingly restrict attention to the first ($-$) one; the other can be
handled analogously.

It is useful to make the unitary transforma\-tion
$U:\hold \ra \hnew$ given by $(U\psi)(y) \linebreak =
e^{-y/2}\psi(e^{-y})$, corresponding to the classical rescaling $R =
e^{-y}.$ Then (\ref{eq:qh}) becomes
\begin{equation}
\hat H_- = -\frac{i\hbar}{\sqrt{12}}\frac{d}{dy},
\label{eq:qhnew}
\end{equation}

\noi and the quantum evolution is
\[\psi(y,t) = \psi\big(y - t/\sqrt{12}\,\big).\]

\noi Thus evolving states in the first summand represent purely
contracting quantum universes; initial states do not split up into
peaks corresponding to contracting and expanding universes as happens,
e.g., in \cite{gd} and \cite{bi}. This is {\em
not\/} because the Hilbert space is an orthogonal direct
sum\footnote{That the first summand arises from quantizing the
component of phase space corresponding to classically contracting
models does not {\em ab initio\/} mean that evolving states belonging
to this summand will contract quantum mechanically. Indeed, this is
what we wish to determine.}; rather, it reflects the fact that
Hamiltonian (\ref{eq:qhnew}) is a first order differential operator.
(Compare
\cite{gd} and \cite{bi}, where the evolution was generated by second
order Hamiltonians.)

To check for quantum collapse, we compute the expectation value of the
radius operator
$\hat R = e^{-y}$ on $\hnew$. Evidently
\[\langle\, \hat R\, \rangle_t = e^{-t/\sqrt{12}}\langle \,\hat R
\,\rangle_0,\]

\noi whence $\langle \,\hat R\, \rangle_t \ra 0$ as $t \ra \infty.$
Thus any initial state $\psi$ with a well-defined radius
must collapse, at least in the sense that
$\lim_{t \ra \infty}\langle\,\hat R\, \rangle_t = 0$.

On the other hand, this quantization is to a certain extent defective,
due to the non-positivity of $\hat H$. (Compare the classical
Hamiltonian (\ref{eq:ch}), which is a positive
function.)\footnote{This phenomenon can be traced to the geometric
fact that the vertical polarization is not complete, cf. \cite{gi2}.}
In an effort to get around this, in \cite{l96}  the
positive square root of the positive self-adjoint operator
\[ {\hat H}^2 = -\frac{\hbar^2}{12}\frac{d^2}{dy^2}.
\label{eq:h2}
\]

\noi was used as the quantum Ham\-iltonian instead of (\ref{eq:qhnew}).
With this modification a ``contradiction'' to Conjecture (F) was
derived, by exhibiting a wave packet which does not collapse as $t \ra
\infty$ relative to the one-parameter unitary group $V_t$ generated by
this operator. Specifically, it was shown that for an initial Gaussian
$\psi,$
\begin{equation}
\lim_{t \ra \infty }\langle V_t \psi |\,\hat R\,|V_t\psi\rangle
\geq \frac{1}{2}\langle \,\hat R
\,\rangle_0.
\label{eq:lev}
\end{equation}

\noi But this result is physically flawed: Indeed, (\ref{eq:lev})
prohibits any (Gaussian) universe from contracting to less than half
its initial radius, even when $\langle \,\hat R
\,\rangle_0$ is arbitrarily large, i.e., for a very classical initial
state. Since under such circumstances quantum effects should be
negligible, it follows that the evolution generated by
$V_t$ does not have the correct classical limit.

Thus $\surd{\hat H^2}$ cannot serve as the quantum Hamiltonian (which,
in any case, is given unambiguously by (\ref{eq:qhnew})), and this
renders the assertions in \cite{l96}
invalid. In fact, we have already proven that relative to the genuine
quantum Hamiltonian, Conjecture (F) {\em is\/} true.

If one regards positivity is a kinematical requirement that must be
preserved by the quantization procedure, then the proper way to
circumvent this problem is to construct an alternate quantization in
such a manner that the resulting Hamiltonian operator is positive.
This we do in the next section. Of course, this alternate quantization
cannot be unitarily equivalent to the quantization presented above.
Regardless, it is a  rigorous quantization for which, as we will show,
the collapse Conjecture (F) is {\em still\/} valid.

\subsection{An Alternate Quantization}

We make the canonical change of coordinates
\[q = R\,\!\pi_{\! R},\;\;p = {\textstyle \frac{1}{2}}\log
\big(|\pi_{\! R}|/\!R\big),\]

\noi so that the phase space becomes the union of $(-\infty,0) \times
\r$ with $(0,\infty) \times \r.$ We quantize in the $q$-coordinate
representation, whence the Hilbert space is $L^2(-\infty,0)
\oplus
\hold \cong \hnew$ with respect to Lebesgue measure
$dq.$ (Note, by means of this isomorphism, that we ``simultaneously''
quantize classically contracting and expanding models.) Then

\begin{displaymath}
\hat H = \frac{1}{\sqrt{12}}\, |\,q\,|
\end{displaymath}

\noi is self-adjoint (on the appropriate domain) and manifestly
positive. The resulting quantum evolution is given by
\[\psi(q,t) = e^{-i|q| t/\sqrt{12}\,\hbar}\psi({q}).\]

To check for quantum collapse, we consider the classical observable $f
= 2p - \log |\,q\,| = - 2\log R,$ since $f \ra \infty$ iff $R \ra 0.$
Then

\[\hat f = -\left(2i\hbar\,\frac{d}{dq} + \log |\,q\,|\right)\]

\noi is essentially self-adjoint on the domain consisting of smooth
functions compactly supported away from zero. As before, it suffices to
consider the dynamics in each summand separately. For an initial state
$\psi$ with $\mbox {supp }\psi \subset (-\infty,0)$, we calculate
\[\langle\, \hat f\, \rangle_t = \langle \,\hat f \,\rangle_0 +
\frac{2t}{\sqrt{12}}.\]

\noi  This implies that
$\langle \,\hat f \,\rangle_t \ra \infty$ as $t \ra \infty$.
Similarly, $\langle \,\hat f \,\rangle_t \ra \infty$ as
$t \ra -\infty$ for an initial $\psi$ with $\mbox {supp }\psi \subset
(0,\infty)$. Thus in this quantization the quantum collapse
Conjecture (F) is also valid.

\section{FRW MODELS}

Next we consider a dust-filled Friedmann-Rob\-ert\-son-Walker (``FRW'')
cosmology. We refer the reader to \cite{gd,l91,lu} for background on
this model.

Prior to an ADM reduction, canonical coordinates are $(R,\pi_{\! R})$
and $(\varphi,\pi_{\varphi})$, where
$\varphi$ is the only nonzero Seliger-Whitham-Schutz velocity
potential for dust and $\pi_{\varphi} > 0.$ The superHamiltonian
constraint is
\begin{equation}
\pi_{\varphi} - \frac{{\pi_{\! R}}^{2}}{24R} -6kR = 0.
\label{eq:sh}
\end{equation}

Now suppose $k = -1$. (The analysis for $k=0$ is similar.) Since
$\pi_{\varphi} > 0$, (\ref{eq:sh}) requires ${|\,\!\pi_{\! R}\,\!| >
12R}$. When
$\pi_{\! R} > 12R$ the model expands from an initial singularity, and
when $\pi_{\! R} < -12R$ it collapses to a final one.

Choosing the slow time
$t = \pi_{\! R}$ and performing an ADM reduction, the Hamiltonian is

\[R(\varphi,\pi_{\varphi},t) = {\displaystyle
\frac{1}{12}}\left(\sqrt{\pi_{\varphi}^2 + t^2} -
\pi_{\varphi}\right).\]

\noi In \cite{l91} this model is canonically quantized in the momentum
representation and it is shown that $\hat R(0)= 0$. {}From this it is
concluded that the $k = -1$ FRW model is singular in this slow-time
gauge, despite the fact that
$\hat R(t)$ is self-adjoint for all $t$.

However, the gauge $t = \pi_{\! R}$, while acceptable classically when
suitably restricted, is not permissible quantum mechanically. The
catch is that the reduced phase space is again disconnected into two
components, corresponding to whether $\pi_{\! R} > 12R$ or $\pi_{\! R}
< -12R$. Fixing one of these components (say the second, corresponding
to a collapsing universe), we see that $\pi_{\! R}$ is {\em a priori}
bounded above by zero. Thus a classical model with
$\pi_{\! R} < 0$ initially can never evolve to a state for which
$\pi_{\! R} > 0$, since this entails collapsing through an infinite
density singularity where one necessarily loses all predictive power.
In fact, for such a model, it does not even make sense to speak of
$\pi_{\! R} > 0.$ But quantum mechanically\footnote{Again see
\cite[\S6.4.3]{ta}.} the Hamiltonian
$\hat R(t)$ is self-adjoint for all times, so that the  evolution is
defined for {\em all\/} $t$. The quantized model therefore transits
through the classical singularity and emerges into an expansion phase.
While the quantized model certainly cannot be said to avoid the
singularity at
$R=0$ (as $\hat R(0) = 0$), it suffers no apparent ``damage''---there
is no loss of predictability, or other pathology---as it collapses.
Thus it is  unclear as to whether the quantum model is actually
singular in any sense.

To avoid this sort of conundrum, in \cite{gd} such choices of time
were specifically excluded. (According to the terminology there, the
gauge $t = \pi_{\! R}$ is not ``dynamically admissible.'') The sine
qua non is that if the time variable does not range over {\em all\/}
of $(-\infty,\infty)$ classically,\footnote{Despite the possibility
that any {\em given\/} classical state may evolve to a singularity in
finite time.} then it is unreasonable to expect it to do so quantum
mechanically, which is what the self-adjointness of the Hamiltonian
requires. Since it is explicitly stated as part of our conjectures in
\cite[p. 2404]{gd} that the choice of time must be dynamically
admissible, these models do not qualify as counterexamples to
Conjecture (S).

We remark that there do exist dynamically admissible
slow-time gauges for the FRW models, e.g., $t=-\varphi.$ The
corresponding quantized models have been shown to be nonsingular
\cite{gd,lu,lr,t}, in accordance with Conjecture (S).

In \cite{l90}, a similar analysis of a $k = 0$ model in the slow-time
gauge $t = \pi_{\mu}$, where $\mu = \log R$, is performed. This choice
of time is likewise dynamically inadmissible.

\section{CONCLUSIONS}

Our calculations show unequivocally that the $k = 0$ \rw\ models
collapse quantum mechanically, at least according to the (standard)
criteria we have employed \cite{gd,lu}. Of course, which of the two
inequivalent quantizations presented in \S2 is physically correct (if
either!) is a matter for speculation. But an essential point is that
whether the quantized models collapse does not depend upon the
positivity of the quantum Hamiltonian. In this regard, we contend that
the alternate quantization in \S2.2 is a completely rigorous and
systematic way of dealing with problems like non-positivity, as
opposed to making ad hoc modifications to the quantum dynamics. We
also draw attention to the treatment of the subtleties arising from the
disconnectedness of the classical phase space in both the RW$\!\phi$
and FRW models.

It will likely require an analysis of more realistic cosmologies to
make sense of the behavior of the quantized FRW models. Regardless,
these examples illustrate the difficulties that arise when the
classical choice of time is {\em ab initio\/} incompatible with the
requirement that the quantum dynamics be generated by a self-adjoint
Hamiltonian.

For an entirely different approach to these issues, we refer the reader to
the recent paper \cite{ab-pn}.


\begin{thebibliography}{9}
\bibitem{gd} Gotay, M.J. and Demaret, J. [1983]  Quantum cosmological
singularities. {\sl Phys. Rev.} {\bf D28}, 2402--2413.

\bibitem{l90} Lemos, N.A. [1990] Inequivalence of unitarity and
self-adjointness: An example in quantum cosmology. {\sl Phys. Rev.}
{\bf D41}, 1358--1359.

\bibitem{l91} Lemos, N. [1991] Singular self-adjoint quant\-um
cosmological models in a slow-time gauge. {\sl Class. Quantum
Grav.} {\bf 8}, 1303--1310.

\bibitem{l96} Lemos, N.A. [1996] Singularities in a scalar field
quantum cosmology. {\sl Phys. Rev.} {\bf D53}, 4275--4279.

\bibitem{bi} Blyth, W.F. and Isham, C.J. [1975] Quantization of a
Friedmann universe filled with a scalar field. {\sl Phys. Rev.} {\bf
D11}, 768--778.

\bibitem{gi1} Gotay, M.J. and Isenberg, J.A. [1980] Geometric
quantization and gravitational collapse. {\sl Phys. Rev.} {\bf D22},
235--248.

\bibitem{gi2} Gotay, M.J. and Isenberg, J.A. [1980]  Can quantum
effects prevent spacetime collapse? In: {\sl Group Theoretical Methods
in Physics}.    Wolf, K.B., Ed. Springer
Lect. Notes in Phys\-ics {\bf
135}, 418-423.

\bibitem{ta} Tate, R.S. [1993] An algebraic approach to the quan\-tization
of constrained systems: Finite  dimensional systems. (Ph.D.
dissertation, Syr\-a\-cuse University, 8/92). Pre\-print gr-qc/9304043.

\bibitem{lu} Lund, F. [1973] Canonical quantization of relativistic
balls of dust. {\sl Phys. Rev.} {\bf D8}, 3253--3259.

\bibitem{lr} Lapchinskii, V.G. and Rubakov, V.A. [1977] Quantum
gravitation: Quantization of the Friedmann model. {\sl Theor. Math.
Phys.} {\bf 33}, 1076--1084.

\bibitem{t} Tipler, F.J. [1987] The structure of the classical
cosmological singularity. In: {\sl Origin and Early History of the
Universe.}  Demaret, J., Ed. Institut d'Astrophysique de l'Universit\'e
de Li\`ege, 339--361.

\bibitem{ab-pn} Acacio de Barros, J. and Pinto-Neto, N. [1996] The
classical interpretation of quantum mechanics and the singularity problem
in quantum cosmology. Preprint gr-qc/9611028.

\end{thebibliography}
\end{document}